\documentclass[12pt]{article}

\usepackage{mathtools}
\mathtoolsset{showonlyrefs}
\usepackage{geometry}
\usepackage{booktabs}
\usepackage{multirow}

\begin{document}

\title{Single-stage, three-arm, adaptive test strategies for non-inferiority trials with an unstable reference}
\author{Werner Brannath, Martin Scharpenberg, Sylvia Schmidt}
%
%
%
%
%
%
%
%

\maketitle

\abstract 
{For indications where only unstable reference treatments are available and use of placebo is ethically justified, three-arm `gold standard' designs with an experimental, reference and placebo arm 
are recommended for non-inferiority trials. In such designs, the demonstration of efficacy of the reference or experimental treatment is a requirement. They have the disadvantage that only little can be concluded from the trial if the reference fails to be efficacious. To overcome this, we investigate novel single-stage, adaptive test strategies where non-inferiority is tested only if the reference shows sufficient efficacy and otherwise $\delta$-superiority of the experimental treatment over placebo is tested. With a properly chosen superiority margin, $\delta$-superiority indirectly shows non-inferiority. We optimize the sample size for several decision rules and find that the natural, data driven test strategy, which tests non-inferiority if the reference's efficacy test is significant, leads to the smallest overall and placebo sample sizes. We proof that under specific constraints on the sample sizes, this procedure controls the family-wise error rate. All optimal sample sizes are found to meet this constraint.
We finally show how to account for a relevant placebo drop-out rate in an efficient way and apply the new test strategy to a real life data set.}

\section{Introduction}
Trial designs with active reference and placebo groups are considered preferable to two-armed non-inferiority designs whenever it is ethical to include the placebo \cite{CPMP/ICH/364/96, EMEA/CPMP/EWP/2158/99,EMA2010,EMA2001}. The use of a placebo group in non-inferiority trials is particularly useful (and easier to justify) in trials where the placebo effect is expected to be of relevance or it is difficult to identify the population which benefits from the reference treatment. It has therefore been recommended in specific indications like asthma, migraine, depression and panic disorder \cite{CPMP/EWP/2922/01,CPMP/EWP/4280/02,CPMP/EWP/518/97,CPMP/EWP/788/01}. If it is ethically justified, a~placebo ($P$) group is included additionally to the groups of the experimental treatment ($E$) and reference treatment ($R$). Frequently, non-inferiority of the new therapy to the reference is sufficient to show, for example if the new therapy has less strong side effects or affects different subgroups of patients. 
However, non-inferiority can easily be shown if neither the experimental nor the reference therapy has an effect in the study. If the reference is known to be unstable in its effect, then a placebo group is needed to show `assay sensitivity' of the study, i.e.\ that it is planned and conducted in a way that an active treatment, like the reference, will be effective. Koch and R\"ohmel\cite{KR05} propose the following hierarchical procedure, which is known as the `gold standard' design for three-armed non-inferiority studies ($\mu_i$ denotes the mean of the effect parameter in group $i\in\{E,R,P\}$, larger means are indicative of more efficacy):
\begin{enumerate}
\item Reject $H_{EP}^S: \mu_E - \mu_P \le 0$, i.e., show superiority of $E$ versus $P$.
\item Reject $H_{ER}^N: \mu_E - \mu_R \le -\delta_N$, i.e., show non-inferiority of $E$ versus $R$ with non-inferiority margin $\delta_N$.
\end{enumerate}
In our examples below, we choose the non-inferiority margin~$\delta_N$ to be half the historically observed effect of the reference over placebo. 

\begin{figure}
\centering
\includegraphics{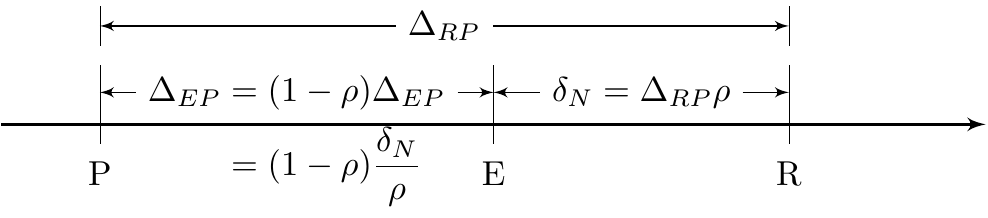}
\caption{\label{fig_deltarho} Illustration of $\delta_N$-non-inferiority of E vs.\ R as a result of $\delta$-superiority of E vs.\ P.}
\end{figure}

As in Hauschke and Pigeot\cite{HP05} one may argue that superiority of the experimental treatment over placebo is not a sufficient conclusion if a reference treatment exists. As a more relevant null hypothesis they introduce the~hypothesis of $\delta$-superiority:
$$
H_{EP}^{\delta}: \mu_E - \mu_P \le \delta.
$$
One may further argue that the true effect of the reference in the study should decide which of the two hypotheses $H_{EP}^{\delta}$ or $H_{ER}^N$ is more meaningful to reject after superiority of the new treatment over placebo has been shown. If the reference's effect is large, then non-inferiority is a~strong assertion, which stands for a successful study (i.e. providing evidence about the efficacy of the new treatment). But if the reference is weak, then there is more interest to show $\delta$-superiority of the new treatment, which is a~success as well. Moreover, if the non-inferiority margin $\delta_N$ is chosen as the fraction $\rho$ of the historical reference effect (i.e.\ the historical difference between the reference and placebo is $\delta_N/\rho$), then $\delta:=(1-\rho)\delta_N/\rho$ is the effect of the experimental treatment group that is equivalent to $\delta_N$-non-inferiority. Hence, $\delta$-superiority of the experimental treatment can be taken as (indirect) proof for non-inferiority, also when the reference fails in the given trial. For an illustration see Figure \ref{fig_deltarho}.

This argumentation seems to imply a~necessity to test the effect of the reference, in order to obtain a~reasonable interpretation of the study. However, the sponsor of the study is basically interested in proving the efficacy of the new treatment and does not wish to spend level in showing efficacy of the standard treatment. In this article, we propose an adaptive extension of the Koch-R\"ohmel design, where the reference effect is not tested but used for a data-driven decision rule (or  
filter) to decide whether to test for non-inferiority or $\delta$-superiority. As we will see in the following, the above idea of interpretation of study success in dependence of the strength of the reference can be implemented formally via a~hierarchical testing procedure that controls the family-wise error rate (FWER), i.e., the probability to make at least one erroneous rejection. Sample size calculation and optimization is done with regard to the considered definition of study success. Even though there exists extensive literature and recent research on three-arm non-inferiority trials,\cite{Muetze17a,Muetze17b,Wu18,Zhong18,Lu18,Xu18,Homma19,Homma21}
we are not aware of a similar suggestion in the literature.

The article is organized as follows. In the next section, we introduce the adaptive testing strategy for three-arm trials. The following section considers different rules for deciding upon the strength of the reference, later also called ``filter'' for the non-inferiority test. It turns out that the intuitive filter requesting a~statistical proof of superiority of the reference over placebo yields the smallest sample sizes needed to obtain a~certain probability of study success. In further two sections, we adjust our optimal sample size calculation, allowing for uncertainty in the true effect of the reference in the study, and for a relevant dropout rate in the placebo group, respectively. We conclude with a~discussion.

\section{Description of the test design} \label{sec_design}

\begin{figure}
\centering
\includegraphics{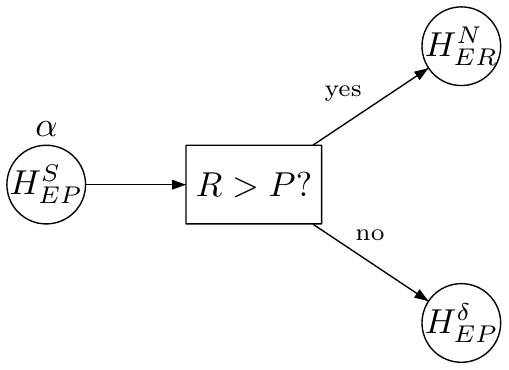}\qquad\qquad
\includegraphics{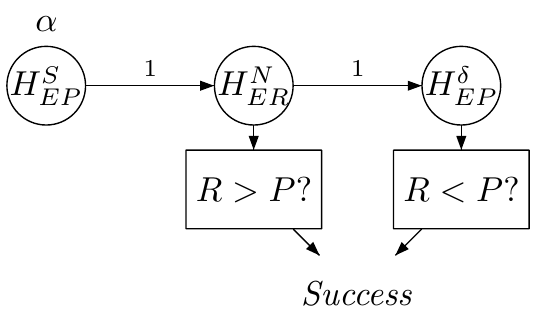}
\caption{\label{fig_design} Left: Intuitive design for the adaptive non-inferiority test strategy. Right: Formal definition of the new design using the graphical notation of Bretz et al.\cite{B09}. The boxes ``$R > P$?" resp. ``$R < P$?" are not tested formally, but serve as rule for the interpretation whether the reference is stronger than placebo or not. See the text for several options how to define this rule.}
\end{figure}

To demonstrate successfully the efficacy of the new treatment, one would like to proceed as sketched in the left graph of Figure~\ref{fig_design}. After showing superiority over placebo, the next hypothesis of interest depends on the strength of the reference. If the reference is strong, i.e., its effect is larger than that of placebo, then one wants to show non-inferiority of the experimental treatment compared to the reference, which is in the spirit of the gold standard design proposed by Koch and R\"ohmel\cite{KR05}. But if the reference is not better than placebo, then non-inferiority to reference is not of interest. Rather, one wants to prove $\delta$-superiority to placebo, which indirectly proves the non-inferiority to the historical reference. Note that the decision ``$R>P$'' serves as a~\textit{filter} for further testing and is not part of the test strategy. Several options are possible to define what ``$R>P$'' means (see below).

The intuitive strategy described above does not guarantee that the FWER is controlled. Although only one of the hypotheses $H_{ER}^N$ or $H_{EP}^\delta$ is tested at level~$\alpha$ after rejection of $H_{EP}^S$, the decision, which of the two hypotheses is tested, is data-driven. Therefore, as a~formal implementation of the intuitive idea, we define the following hierarchical test, which is shown in the right graph of Figure~\ref{fig_design}. This figure uses the notation of graphical test procedures introduced in Bretz et al.\cite{B09} and includes the filter as exploratory test to interpret the result. After rejection of $H_{EP}^S$, we pass the level~$\alpha$ to the non-inferiority hypothesis $H_{ER}^N$, in concordance with the gold standard design. But in contrast to the proposed continuations in the Koch-R\"ohmel design, after rejection of the non-inferiority hypothesis, the next hypothesis in our hierarchical sequence to be tested at level~$\alpha$ is the $\delta$-superiority $H_{EP}^\delta$. The filter ``$R>P$'' now serves as interpretation for the \textit{success} of the study. Success means that we have evidence about the efficacy of the new treatment. This can either be reached if the reference is strong (i.e., the filter is satisfied) and non-inferiority of the new treatment over the reference is shown, or if the reference is weak (i.e., the filter is not satisfied) and $\delta$-superiority of the new treatment is shown. Since $\delta$ is $\rho$ times the historically observed reference effect, this serves as an indirect proof of non-inferiority over the (historical) reference.

For application, the left side of Figure~\ref{fig_design} would be easier to explain to practitioners. The hierarchical strategy in the right side of Figure~\ref{fig_design} seems unnatural, because one tests first $E$ versus~$P$, then $E$ versus~$R$, and then again $E$ versus~$P$. Assume that we apply the intuitive strategy using the following filter: The reference is declared ``strong'' if its superiority is shown at significance level~$\alpha$, i.e., if
\begin{equation} \label{filter1}
\frac{X_R-X_P}{\sigma\sqrt{n_R^{-1}+n_P^{-1}}} \ge z_\alpha,
\end{equation}
where	$X_R$ and $X_P$ are the observed group means in the reference and placebo group, $\sigma$ is the common standard deviation, $n_R$ and $n_P$ are the sample sizes in the two groups and $z_\alpha$ is the $(1-\alpha)$-quantile of the normal distribution.	We will show in the Appendix that, in a~large number of situations, the intuitive strategy with this filter leads to the same decisions concerning success of the study as the formal strategy with the same filter. This seems curious, because the formal strategy requires proof of non-inferiority also in the case, where only $\delta$-superiority of $E$ versus~$P$ is of interest, which should reduce the success probability. However, $\delta$-superiority of interest only if the reference is not better than placebo, i.e.\ $X_R\le X_P$, so that non-inferiority of an effective new treatment is not a~high barrier to take.

The equivalence between the two strategies is implied if a~certain restriction on the sample sizes is satisfied. A~simple check shows that this restriction holds for all optimal sample sizes derived in the remainder of this article. Hence, the new design can be interpreted in an intuitive way, while assuring strong error control. We call this new design \textit{single-stage adaptive non-inferiority design for three-arm trials}.

\section{Filter options} 

The filter in~\eqref{filter1} seems to be a~natural condition for deciding that we have a strong reference, i.e., to evidence that ``$R>P$'' holds in the two graphs of Figure~\ref{fig_design}. There are some more natural candidates for indicating that the reference has a~relevant effect over placebo. For simplicity, we will from now on assume that $\rho=1/2$, i.e.\ $\delta=\delta_N$ are equal to half of the historical reference effect. We will discuss the following four options (with notations as above):
\begin{itemize}
\item Filter 1: ${(X_R-X_P)}\Big/{\sigma\sqrt{n_R^{-1}+n_P^{-1}}} \ge z_\alpha$, i.e., superiority as in~equation~\eqref{filter1}
\item Filter 2: ${(X_R-X_P-\delta)}\Big/{\sigma\sqrt{n_R^{-1}+n_P^{-1}}} \ge z_\alpha$, i.e., superiority by half the historical reference effect is shown at significance level~$\alpha$
\item Filter 3: $X_R-X_P \ge 2\delta$, i.e., the observed difference of means is at least the historical difference
\item Filter 4: $X_R-X_P \ge 3/2 \delta$, i.e., the observed difference of means is at least $3/4$ of the historical difference
\end{itemize}

To evaluate the filter options, we consider the required sample sizes to obtain a~probability of~$90\%$ for the success of the study, i.e., to show either non-inferiority of the new treatment to the reference in the case where the filter is satisfied (denoted by ``Power~1'' in the following), or to show $\delta$-superiority of the new treatment over placebo in the case where the filter is not satisfied (denoted by ``Power~2''). The overall success probability (``Total Power'') is the sum of Power~1 and Power~2.

Schl\"omer and Brannath\cite{SB13} have developed an algorithm, which minimizes the total sample size that is needed to obtain a~pre-specified power for the original gold standard design, i.e. to show non-inferiority of the new treatment compared to the reference. By adjusting this algorithm, we can calculate optimal sample sizes to reach a~desired power for the success of the study. We make the following assumptions for the calculation:
\begin{itemize}
\item desired power $90\%$ for the success of the study
\item one-sided FWER $\alpha=2.5\%$
\item independent normal observations with common standard deviation $\sigma=0.5$
\item non-inferiority margin $\delta=0.1$, which corresponds to half of the historical reference effect
\item the effect of the new treatment is equal to the historical reference effect, $\mu_{E} - \mu_P=0.2$
\end{itemize}

We consider three different scenarios for the effect of the reference over the placebo:
\begin{itemize}
\item Scenario 1: the reference is as good as observed historically, $\mu_{R} - \mu_P = 0.2$
\item Scenario 2: the reference is only half as good as observed historically, $\mu_{R} - \mu_P=0.1$
\item Scenario 3: the reference fails and is as good as placebo, $\mu_{R} - \mu_P=0$
\end{itemize}


\begin{table}
\centering
\caption{\label{tab_filter} Optimal sample sizes for the flexible non-inferiority design with different filter options and for different scenarios (see text for details)}
\begin{tabular}{cccccc}				
\toprule
	Scenario &&  Filter 1 & Filter 2 & Filter 3 & Filter 4 \\
\midrule
 \multirow{4}{5pt}{1} & $n_E$ &	538	& 608 & 741	& 611 \\
& $n_R$ &	547	& 610 & 548 & 607 \\
 & $n_P$ &	159 & 458 &	547 & 366 \\\cmidrule{2-6}
& $N$ &	1244  & 1676 &	1836  &	1584  \\
\midrule
\multirow{4}{5pt}{2} & $n_E$ &	288	& 546 & 546	& 542 \\
 & $n_R$ &	284	& 143 & 143 & 145 \\
 & $n_P$ &	472 & 533 &	533 & 530 \\\cmidrule{2-6}
& $N$ &	1044  & 1222 &	1222  &	1217  \\
\midrule
\multirow{4}{5pt}{3} & $n_E$ &	531	& 532 & 532	& 532 \\
 & $n_R$ &	68	& 67 & 67 & 67 \\
 & $n_P$ &	529 & 529 &	529 & 529 \\\cmidrule{2-6}
& $N$ &	1128  & 1128 &	1128  &	1128  \\
\bottomrule
\end{tabular}
\end{table}

Table~\ref{tab_filter} shows the calculated optimal sample sizes for each of the three scenarios and every filter. The most intuitive filter~1, which requires a~statistical proof at $\alpha$-level of superiority of the reference over placebo, leads to the smallest sample sizes in each scenario. In particular, as long as the reference is effective such that non-inferiority suffices for the success of the study, the size of the placebo group is small compared to the other groups, which is ethically favorable. We will therefore in the following consider only the first option for the filter, as given by~\eqref{filter1}.

%
Comparing the scenarios, the highest total sample size is needed with all filters in Scenario~1 and the lowest sample size in Scenario~3. In the latter scenario, there is almost no difference between the filters concerning sample size allocation. It is clear that if the reference effect is zero, then success is mainly reached by showing $\delta$-superiority of the experimental treatment over placebo. Hence, the reference sample size is very small with all filters.

When considering Scenario~2, we observe that filters 2-4 all present with almost identical optimal sample sizes. This is due to the fact that in this setup where the reference has only half the historic effect, these filters indicate that ``$R>P$'' with almost equal probability (in fact they conclude ``$R>P$'' with very small probability). As a consequence, these three filters lead to almost the same testing procedure, needing to show $\delta$-superiority of the experimental treatment over placebo to claim success.

\section{Accounting for uncertainty in the reference effect} \label{sec_reference}

As can be seen from Table~\ref{tab_filter}, sample sizes are quite sensitive to the assumed scenario. This is also illustrated in Table~\ref{tab_power12}, where the contributions of Power~1 (from non-inferiority) and Power~2 (from $\delta$-superiority) to the total power in the three scenarios considered above are indicated: If the reference effect corresponds to the observed historical effect (Scenario~1), then all power is gained by the non-inferiority claim. In the second scenario, superiority of the reference fails to be shown in almost $25\%$ of the cases, therefore some power ($14\%$) must be gained by showing $\delta$-superiority of the new treatment over placebo. If the reference fails completely, then almost all power is gained by the $\delta$-superiority.

Another issue is the question how sensitive the power behaves if the true scenario is actually different from the made assumptions. So far, we presumed that the true scenario is the same as the one assumed for the sample size calculation. Table~\ref{tab_scenarios} shows, for the three considered scenarios, that substantial power losses may occur when the assumptions made do not hold. This is comprehensible because the optimal sample size allocations between the three groups rely on either of the two different tests to reach the desired power. If the reference effect is weaker or stronger than assumed, then the respective other test is successful, which has not been powered for.

\begin{table}
\centering
\begin{minipage}[t]{0.5\textwidth}
\caption{\label{tab_power12} Different power strategies with optimal sample sizes in the three scenarios\newline}
\begin{tabular}{lcccc}
\toprule
(Filter 1) &&	Sc 1 &	Sc 2 &	Sc 3 \\
\midrule
Filter satisfied && 	99.3\% &	75.9\% &	2.5\% \\
Power1 &&	90.0\%	& 75.6\% &	2.2\% \\
Power2 &&	0.0\%	& 14.4\% &	87.8\% \\
\midrule
Total Power &&	90.0\% &	90.0\% &	90.0\% \\
\bottomrule
\end{tabular}

\end{minipage}\hfill%
\begin{minipage}[t]{0.46\textwidth}
\caption{\label{tab_scenarios} Change of power if true scenario is different from the scenario for which the sample size was calculated}
\begin{tabular}{lccccc}
\toprule
 (Filter 1)      &&&	\multicolumn{3}{c}{Power in Scenario} \\		
                 &&&		1 &	2 &	3 \\
\midrule
Sample size   &	1	&& 90.0\% &	70.8\% &	60.1\% \\
 for Scenario &	2 &&	66.6\%	 & 90.0\%  & 	76.3\% \\
              &	3 && 32.3\% &	80.0\% &	90.0\% \\
\bottomrule
\end{tabular}
\end{minipage}
\end{table}

Therefore, we propose an alternative sample size calculation. We consider several possible ratios~$v$ of the reference effect in the study divided by the historical reference effect. Each option for~$v$ is weighted by an assumed probability of its occurrence. More general, denote the density of the probability distribution of $v$ on $[0,1]$ with $f(v)$. Then the weighted success probability is given by 
\begin{equation} \label{success_general}
S = \int_0^1 S_{v}f(v)dv,
\end{equation}
where $S$ is the targeted success probability and $S_v$~is the success probability if $v$ is the true ratio. 

A simple and plausible choice for the distribution of $v$ can be derived as follows: Since we do not expect the reference to fail completely, we expect $v=1$ with some probability~$p$ and we assume with probability $(1-p)/2$ that $v=3/4$ and $v=1/2$, respectively. The success probability in \eqref{success_general}, which is targeted at~$90\%$, as before, is then given by
\begin{equation} \label{success}
S = p S_{1} + (1-p)/2 S_{3/4} + (1-p)/2 S_{1/2}.
\end{equation}

\begin{figure}
\includegraphics[width=\textwidth]{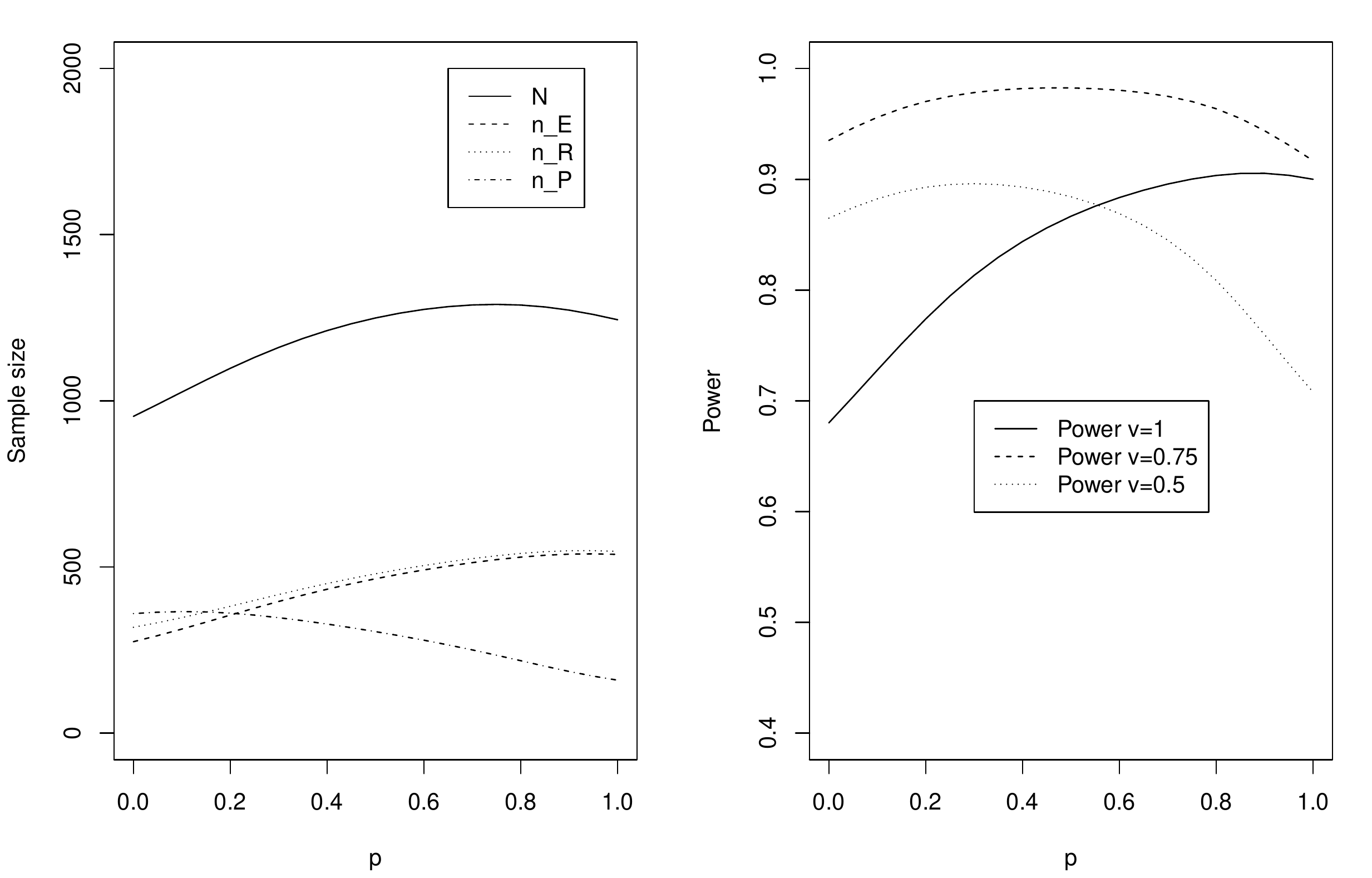}
\caption{\label{fig_p} Optimal sample sizes for the three groups and total (left), and probability of success (right) for different true ratios $v=$ (reference effect in study)/(historical effect), when sample sizes were calculated with different probabilities~$p$ for $v=1$ ($x$-axis).}
\end{figure}

Adjusting the algorithm of Schl\"omer and Brannath\cite{SB13} accordingly (with all other parameters as in the previous section), we obtain new optimal sample sizes for the three groups. Figure~\ref{fig_p} (left) illustrates the change of sample sizes for different values of~$p$. In Figure~\ref{fig_p} (right), the success probabilities are shown in dependence of~$p$ if the true reference effect is $v=1$, $3/4$ or $1/2$ of the historical effect, respectively.

One sees from Figure~\ref{fig_p} that the overall sample size is not monotone in~$p$, the maximum being reached for $p=0.75$. The larger $p$ is, i.e., the more likely one expects the reference to be strong, the lower is the optimal sample size for the placebo group. Also, the power shows no monotonicity in~$p$ and likewise not in the true ratio~$v$. The highest power is gained for all~$p$ when the true reference effect is $v=3/4$. Without any knowledge of the expected performance of the reference, a~probability~$p$ between $0.5$ and $0.6$ is recommended to be used for the sample size calculation to avoid large power losses in the case of an~extremely strong or extremely weak reference.

\section{Modification to account for ethical considerations} \label{sec_placebo}

In general, one wants to have a~small placebo group in order to offer an effective treatment to most of the patients. The size of the placebo group has also practical relevance, because placebo patients are most likely to leave the study early (even if the study is blinded) so that additional recruitment to the placebo group is necessary. An idea to take the high drop-out rate into account is to include the required extra-recruitment in the optimization procedure. To this end, let $n_P$ be the number of evaluable patients and $w_P n_p$, $w_P>1$, the required number of recruited patients. Instead of minimizing the sample size $n_P+n_R+n_E$, we target now at $w_P n_P+n_R+n_E$, where $w_P>1$. When minimizing this quantity to obtain a~fixed success probability of the study, the optimal size of the placebo group is reduced, which is in favor of ethical considerations. 

We adjusted the algorithm of Schl\"omer und Brannath\cite{SB13}, incorporating both the weighted success probability~\eqref{success} and a drop-out parameter~$w_P$, leaving the other parameters as before. We compare the results from the previous section (where $w_P=1$) to a setting, where half of the recruited patients in the placebo group drop out ($w_P=2$). Table~\ref{tab_penalty} compares sample sizes and power results for different values of~$p$ and~$v$. The probability~$p$ for the scenario that the reference is as strong as historically ($v=1$) is chosen not smaller than $1/2$, because we believe that this is a minimum requirement for the reference in a clinical trial. The case $p=1$ excludes the uncertainty about the reference effect and therefore corresponds to Scenario~1 in the section on the filter options.

When assuming $p=1$ for the sample size calculation, then the sample size of the placebo group is reduced from 159 if $w_P=1$ to 139 if $w_P=2$. Of course, we need to initially recruit $2\times 139=278$ patients, but ignoring the drop-out in the optimization procedure one would need to recruit $2\times 159=318$ patients, i.e. 40 more, to the placebo group. 

On the other hand, the total planned sample size increases from 1244 to 1253. The expected number of recruited patients is $n_E+n_R+2n_P=1392$ in the case where this quantity is minimized, compared to 1403 if we do not take the drop-out rate into account. Furthermore, a~prize has to be paid in terms of scenario uncertainty. If, for example, the reference in the study has only half the historical effect ($v=0.5$), then the probability of a~successful study decreases from $71\%$ if $w_P=1$ to $66\%$ if $w_P=2$.

\begin{table}
\centering
\caption{\label{tab_penalty} Sample sizes and power (probability of success) for different probabilities $p$ of Scenario 1 ($v=1$) and probabilities $(1-p)/2$ for $v=0.75$ resp. $v=0.5$, where $v=$ (reference effect in study)/(historical effect). Results are compared for $w_P=1$ (minimization of planned sample size) and $w_P=2$ (minimization of expected number of recruited patients).}
\begin{tabular}{lcrrcrrcrrc}				
\toprule
 && \multicolumn{2}{c}{$p=1$} &&	\multicolumn{2}{c}{$p=0.8$} && \multicolumn{2}{c}{$p=0.5$} \\				
 && $w_P=1$ &	$w_P=2$ && $w_P=1$ & $w_P=2$ && $w_P=1$ &	$w_P=2$ \\
\midrule
$n_E$ &&	538 &	551 &&	530 &	555 &&	465 &	500 \\			
$n_R$ &&	547 &	563 &&	541 &	572 &&	479 &	524 \\			
$n_P$ &&	159 &	139 &&	218 &	179 &&	305 &	249 \\			
$N$ planned &&	1244	 & 1253 &&	1289 &	1306 &&	1249 &	1273 \\
$N$ recruited &&	1403 &	1392 &&	1507 &	1485 &&	1555 &	1522 \\
\midrule
Power $v=1$ &&	90.0\% &	90.0\% &&	90.3\% &	91.4\% &&	 86.7\% &	89.2\% \\			
Power $v=0.75$ &&	91.7\% & 	88.8\% &&	96.4\% &	94.0\% &&	98.2\% &	97.4\% \\			
Power $v=0.5$  && 70.8\% &	66.0\% &&	80.9\% &	74.9\% &&	88.4\% &	84.2\% \\
\bottomrule
\end{tabular}
\end{table}

\section{Example}
We next apply the proposed testing procedure to data of a three-arm study on major depressive disorder by Higuchi~et al.\cite{Higu11}. We will first show the example in it's original version, as shown in \cite{Higu11}. After that, we will apply the proposed procedure to the same data.  The primary objective of this double-blinded, randomized, active controlled, parallel-group study was to compare the efficacy and safety of 6-week treatment with duloxetine (E) to those of paroxetine (R) and placebo (P).  The primary endpoint of the study was the HAM-D17 change from baseline at 6 weeks and the statistical analysis was planned to test the superiority of duloxetine over placebo and the non-inferiority of duloxetine over paroxetine in hierarchical order. The non-inferiority margin was set as $\delta_N=2.5$. The observed mean decreases were $10.2\pm 6.1$ (mean$\pm$sd) in the duloxetine group, $9.4\pm 6.9$ in the paroxetine group and $8.3\pm 5.8$ in the placebo group with sample sizes $n_E=147,\ n_R=148,\ n_P=145$.
The two-sided 95\% confidence interval for the difference in means between duloxetine and placebo is in this case given by (0.53,\ 3.27), indicating superiority of duloxetine over placebo. The 95\% confidence interval for the difference in means between duloxetine and paroxetine can be calculated to be (-0.69,\ 2.30) and excludes $-\delta_N=-2.5$ indicating non-inferiority of duloxetine compared to paroxetine. However, the superiority of paroxetine over placebo could not be established, because the 95\% confidence interval for this comparison is given by (-0.37,\ 2.57) and therefore includes 0. Higuchi~et al.\cite{Higu11} concluded that non-inferiority of duloxetine compared to paroxetine did not have assay sensitivity.

The trial data were also investigated by Hida and Tango \cite{HT11}, who argue that for the proof of assay sensitivity it does not suffice to show that $\mu_R>\mu_P$ but one rather needs to show that the reference is at least $\delta$-superior to placebo ($\mu_P<\mu_R-\delta$). Since, paroxetine could not be shown to be superior to placebo, it is also not considered to be $\delta$-superior. Therefore, Hida and Tango \cite{HT11} also conclude the lack of assay sensitivity. It should be noted that in the given setup, the requirement of showing $\delta$-superiority of reference over placebo is very similar to filter~2 introduced above. However, in the approach of Hida and Tango \cite{HT11}, the testing procedure stops if $\delta$-superiority can not be established, while in our proposed approach, we would still have the opportunity to obtain a successful study by showing $\delta$-superiority of the experimental treatment over placebo. Following the approach outlined in the introduction (cf. Figure \ref{fig_deltarho}) $\delta$ is chosen in a way that $\delta$-superiority of E over P is equivalent to $\delta_N$-non-inferiority of E to the historical reference effect. In their sample size calculation Higuchi~et al.\cite{Higu11} assume a historical effect of R over P of 5. Therefore, in our application we chose $\delta=\delta_N=2.5$.

\subsection{Application of the proposed method}
We now apply the proposed testing procedure to the data mentioned above. As stated above, duloxetine can be shown to be superior to placebo (i.e.\ $H_{EP}^S$ is rejected). However, using filter~1, ``$R>P$'' can not be concluded, which is why as a next step we test for $\delta$-superiority of duloxetine over placebo ($H_{EP}^\delta$). Since the 95\% confidence interval for this comparison includes $\delta=2.5$, $H_{EP}^\delta$ can not be rejected. Therefore, for the given data, using the newly proposed methods leads to the same conclusion as the methods applied originally. Note that we applied the intuitive testing approach, instead of the formal definition. In the case of this study both approaches are equivalent (because the sample sizes fulfill the special condition derived in the appendix).

For illustrative purposes we assume that the observed mean decrease in HAM-D17 at week 6 in the duloxetine group was $12.2\pm 6.1$ (mean$\pm$sd). We re-calculate the confidence interval for the comparison of the duloxetine and placebo groups as (2.53,\ 5.27). Under this assumption, duloxetine is of course still superior to placebo. Furthermore, using filter~1 ``$R>P$'' can not be concluded, which is why as a next step we test for $\delta$-superiority of duloxetine over placebo, which in this case can be concluded since the aforementioned confidence interval excludes $\delta=2.5$. Hence, the study would be successful in the sense that, the experimental treatment could be shown to be sufficiently effective. Note that the standard approach to the testing of non-inferiority as well as the approach by Hida and Tango \cite{HT11} still conclude the lack of assay sensitivity and stop there, without considering the fact that the experimental treatment is sufficiently effective to claim success even when the reference is weak.

Note, that the confidence intervals reported in this section, while appropriate for deriving the test decision in the hierarchical test, are no simultaneous confidence intervals, and therefore do not have simultaneous coverage probability. If one is interested in informative simultaneous confidence intervals, the procedure of Schmidt and Brannath \cite{SB14} could be implemented. In that case appropriate splitting weights would have to be defined, which is beyond the scope of this paper. Future research on this topic might involve defining these weights in dependence on the filter chosen in our procedure. 

\subsection{Sample size}
For the calculation of the optimal sample size, we use the assumptions proposed by Hida and Tango \cite{HT11}: (i) duloxetine and paroxetine have the same effect size ($\mu_E=\mu_R=10$), (ii) the placebo has half the effect ($\mu_P=5$), (iii) the three arms have a common standard deviation $\sigma=6.5$ and (iv) the same value as in the above-stated study for the non-inferiority margin is chosen: $\delta=2.5$. Under these assumptions, the optimal sample size to achieve a power of 80\% (which is the same as assumed in Hida and Tango\cite{HT11}) for our proposed method at the one-sided significance level $0.025$ is: $n_E=110,\ n_R=114,\ n_P=39$, resulting in an overall sample size of $N=263$ with filter~1 and $n_E=130,\ n_R=131,\ n_P=101$ resulting in an overall sample size of $N=362$ with filter~2. These sample sizes are considerably smaller than those calculated by Hida and Tango\cite{HT11}, who need  $n_E=n_R=151,\ n_P=121\ (N=423)$ under the same assumptions. Furthermore, both sample sizes for the new approach fulfill the condition derived in the appendix, leading to equivalence between the intuitive design and the formal design for the new testing procedure.

\section{Summary and discussion} \label{sec_discussion}
The gold standard design is applied in indications with an unstable reference. Proving efficacy of the reference is not its main goal and only necessary if success of the new treatment is shown via non-inferiority compared to reference. This idea is made explicit with the adaptive testing strategy introduced here. It provides the possibility to assure success also with a~weak reference by showing $\delta$-superiority of the new treatment over placebo.

A~flexibility is proposed also with respect to the uncertainty of the reference performance by underlying a~scenario mixture in the sample size calculation. We propose to assume a~strong or partly strong reference with probability around one half, respectively, yielding high success probabilities for any true reference effect. Furthermore, an adjustment of sample size is considered for the case of a high drop-out rate in the placebo group.

The new design can easily be communicated to practitioners by an intuitive graphic. Here, it has to be taken care that the sample sizes satisfy a~certain condition (see Appendix), which guarantees equivalence of the intuitive picture to a~more formal one and thus control of the family-wise error rate. This was never a~problem in all settings considered here. Further investigation might be done to make this formal condition more comprehensible or maybe to adapt the filter in a~way that no such condition is necessary.

The idea of testing non-inferiority and $\delta$-superiority may alternatively be implemented within a Bayesian framework. Here we could use the joint posterior distribution of the three effect parameters to calculate e.g. the posterior probability that the experimental treatment is non-inferior to the reference and the reference is superior to placebo and the posterior probability that the experimental treatment is $\delta$-superior to placebo. The treatment could then be claimed useful if at least one these two posterior probabilities are sufficiently high. However, since confirmatory claims often rely on frequentist methods, we focused on the frequentist approach.

A further topic of future research might be deriving informative simultaneous confidence intervals in the framework defined in this paper.

\section*{Acknowledgments}
We thank Anna Schritz for her help in the development of the R programs which were used for this manuscript. 

\section*{Declaration of conflicting interests}
The authors declared no potential conflicts of interest with respect to the research, authorship, and/or publication of this article.

\section*{Funding} 
The authors disclosed receipt from the following financial support for the research, authorship, and/or authorship of this article: 
This work was supported by the German Research Foundation (DFG) under grant BR 3737/1-1. 

\bibliographystyle{plain}
\bibliography{Literatur}

\begin{thebibliography}{10}

\bibitem{CPMP/ICH/364/96}
CPMP.
\newblock Ich topic e 10. choice of control group in clinical trials.
\newblock {\em European Medicines Agency, London}, (CPMP/ICH/364/96), 2001.

\bibitem{CPMP/EWP/2922/01}
CPMP.
\newblock Note on guidance on the clinical investigation of medicinal products
  in the treatment of asthma.
\newblock {\em European Medicines Agency, London}, (CPMP/EWP/2922/01), 2002.

\bibitem{CPMP/EWP/518/97}
CPMP.
\newblock Note on guidance on the clinical investigation of medicinal products
  in the treatment of depression.
\newblock {\em The European Agency for the Evaluation of Medical Products,
  London}, (CPMP/EWP/518/97), 2002.

\bibitem{EMEA/CPMP/EWP/2158/99}
CPMP.
\newblock Guideline on the choice of the non-inferiority margin.
\newblock {\em European Medicines Agency, London}, (EMEA/CPMP/EWP/2158/99),
  2005.

\bibitem{CPMP/EWP/4280/02}
CPMP.
\newblock Note on guidance on the clinical investigation of medicinal products
  in the treatment of panic disorder.
\newblock {\em European Medicines Agency, London}, (CPMP/EWP/4280/02), 2005.

\bibitem{CPMP/EWP/788/01}
CPMP.
\newblock Note on guidance on the clinical investigation of medicinal products
  in the treatment of migraine.
\newblock {\em European Medicines Agency, London}, (CPMP/EWP/788/01), 2007.

\bibitem{HP05}
Hauschke D and Pigeot I.
\newblock Establishing efficacy of a new experimental treatment in the `gold
  standard' design.
\newblock {\em Biom J}, 47(6):782--786, 2005.

\bibitem{HT11}
Hida E and Tango T.
\newblock On the three-arm non-inferiority trial including a placebo with a
  prespecified margin.
\newblock {\em Stat Med}, 30:224--231, 2011.

\bibitem{EMA2010}
EMA.
\newblock Reflection paper on the need for active control in therapeutic areas
  where use of placebo is deemed ethical and one or more established medicines
  are available.
\newblock {\em European Medicines Agency, London}, (Draft, EMA/759784/2010),
  2010.

\bibitem{EMA2001}
EMEA.
\newblock Emea/cpmp position statement on the use of placebo in clinical trials
  with regard to the revised declaration of helsinki.
\newblock {\em The European Agency for the Evaluation of Medical Products,
  London}, (EMEA/17424/01), 2001.

\bibitem{B09}
Bretz F, Maurer W, Brannath W, and Posch M.
\newblock A graphical approach to sequentially rejective multiple test
  procedures.
\newblock {\em Stat Med}, 28(4):586--604, 2009.

\bibitem{Homma19}
Homma G and Diamon T.
\newblock Sequential parallel comparison design for “gold standard”
  noninferiority trials with a prespecified margin.
\newblock {\em Biometrical Journal}, 61:1493--1506, 2019.

\bibitem{Homma21}
Homma G and Diamon T.
\newblock Sample size calculation for “gold-standard” noninferiority trials
  with fixed margins and negative binomial endpoints.
\newblock {\em Statistics in Biopharmaceutical Research}, 13:435--447, 2021.

\bibitem{Xu18}
Lu~H, Jin H, and Zeng W.
\newblock A more efficient three-arm non-inferiority test based on pooled
  estimators of the homogeneous variance.
\newblock {\em Stat Methods Med Res}, 27:2437--2446, 2018.

\bibitem{Zhong18}
Zhong J, Wen MJ, Kwong KS, and Cheung SH.
\newblock Testing of non-inferiority and superiority for three-arm clinical
  studies with multiple experimental treatments.
\newblock {\em Stat Methods Med Res}, 27:1751--1765, 2018.

\bibitem{KR05}
A~Koch and R\"ohmel J.
\newblock Hypothesis testing in the ``gold standard'' design for proving the
  efficacy of an experimental treatment relative to placebo and a reference.
\newblock {\em J Biopharm Statist}, 14(2):315--325, 2004.

\bibitem{SB13}
Schl\"omer P and Brannath W.
\newblock Group sequential designs for three-arm `gold standard'
  non-inferiority trials with fixed margin.
\newblock {\em Stat Med}, 32(28):4875--4889, 2013.

\bibitem{SB14}
Schmidt S and Brannath W.
\newblock Informative simultaneous confidence intervals in hierarchical
  testing.
\newblock {\em Methods Inf Med}, 53, 2014.

\bibitem{Higu11}
Higuchi T, Murasaki M, and Kamijima K.
\newblock Clinical evaluation of duloxetine in the treatment of major
  depressive disorder-placebo- and paroxetine-controlled double-blind
  comparative study.
\newblock {\em Jpn J Clin Psychopharmacol}, 12:1613--34, 2009.

\bibitem{Muetze17a}
M\"utze T, Konietschke F, Munk A, and Friede T.
\newblock A studentized permutation test for three-arm trials in the `gold
  standard' design.
\newblock {\em Stat Med}, 36:883--898, 2017.

\bibitem{Muetze17b}
M\"utze T and Friede T.
\newblock Blinded sample size re-estimation in three-arm trials with `gold
  standard' design.
\newblock {\em Stat Med}, 36:3636--3653, 2017.

\bibitem{Lu18}
Xu~W, Hu~F, and Cheung SH.
\newblock Adaptive designs for non-inferiority trials with multiple
  experimental treatments.
\newblock {\em Stat Methods Med Res}, 27:3255--3270, 2018.

\bibitem{Wu18}
Wu~Y, Li~Y, Hou Y, Li~K, and Zhou X.
\newblock Study duration for three-arm non-inferiority survival trials designed
  for accrual by cohorts.
\newblock {\em Stat Methods Med Res}, 27:507--520, 2018.

\end{thebibliography}

\appendix

\section{Comparison of the intuitive and the formal design} \label{sec_appendix}

Our discussion restricts to the preferred Filter 1, other filters can be handled similarly. We use in this section the following notations for the relevant test statistics:
\begin{align*}
T_{EP}^\delta &= \frac{X_E-X_P-\delta}{\sigma\sqrt{n_E^{-1}+n_P^{-1}}}, \quad
T_{RP}^S =\frac{X_R-X_P}{\sigma\sqrt{n_R^{-1}+n_P^{-1}}},\\
T_{ER}^N &=\frac{X_E-X_R + \delta_N}{\sigma\sqrt{n_E^{-1}+n_R^{-1}}}.
\end{align*}
These test statistics are not independent, in fact,
\begin{equation}
\label{trp}
T_{RP}^S = \frac{\sqrt{n_E^{-1}+n_P^{-1}}T_{EP}^\delta - \sqrt{n_E^{-1}+n_R^{-1}}T_{ER}^N+(\delta_N+\delta)/\sigma}{\sqrt{n_R^{-1}+n_P^{-1}}} .
\end{equation}
The only difference between the formal and the intuitive approach is that the latter interprets observations as leading to a~success of the study, where the filter is not satisfied and $\delta$-superiority of $E$ versus~$P$ can be shown, whereas formally it is additionally necessary that $E$ is non-inferior to~$R$ because of the hierarchical structure, which guarantees error control. Hence, different decisions with the two strategies occur if and only if
\begin{equation}\label{tieq}
T_{RP}^S < z_\alpha 
\quad\text{and}\quad
T_{EP}^\delta \ge z_\alpha 
\quad\text{and}\quad
T_{ER}^N < z_\alpha.
\end{equation}
By \eqref{trp}, we have
\begin{align*}
&T_{RP}^S < z_\alpha \\
&\iff \\
&T_{ER}^N  > \frac{\sqrt{n_E^{-1}+n_P^{-1}}T_{EP}^\delta-\sqrt{n_R^{-1}+n_P^{-1}}z_\alpha + (\delta_N+\delta)/\sigma}
								  {\sqrt{n_E^{-1}+n_R^{-1}}} =: B
\end{align*}
Now, the condition $T_{ER}^N < z_\alpha$ implies $B < z_\alpha$, or equivalently,
$$
T_{EP}^\delta < \frac{\left(\sqrt{n_E^{-1}+n_R^{-1}}+\sqrt{n_R^{-1}+n_P^{-1}}\right)z_\alpha - (\delta_N+\delta)/\sigma}
								  {\sqrt{n_E^{-1}+n_P^{-1}}} =: C
$$
Furthermore, $T_{EP}^\delta\ge z_\alpha$ necessitates $z_\alpha < C$. As a consequence, situation \eqref{tieq}
cannot occur when $z_\alpha\ge C$, and then the naive and formal procedures coincide. One easily verifies that
 $z_\alpha\ge C$ is equivalent to 
\begin{equation}
\label{cond}
\left(\sqrt{n_E^{-1}+n_R^{-1}}+\sqrt{n_R^{-1}+n_P^{-1}}-\sqrt{n_E^{-1}+n_P^{-1}}\right)z_\alpha \le (\delta_N+\delta)/\sigma.
\end{equation}
This condition depends only on the sample sizes of the three groups. It is generally satisfied in interesting situations, where basically the reference group is not too small. A~closer look shows that condition~\eqref{cond} is satisfied for all optimal sample sizes derived in this article. Furthermore, we have determined that the optimal sample sizes when applying the intuitive design are exactly the same as for the formal design in all situations discussed here. Hence, it is very often not necessary to distinguish between the two designs so that the new adaptive design will be easier to communicate to practitioners.

\end{document}